\documentclass[prb, aps,
twocolumn,
showpacs,amsmath,amssymb]{revtex4}
\usepackage{epsf}
\usepackage{graphicx}

\usepackage[usenames]{color}

\usepackage[normalem]{ulem}

\def \be {\begin{equation}}
\def \ee {\end{equation}}

\def \bea {\begin{align}}
\def \eea {\end{align}}

\def\bee{\begin{eqnarray}}
\def\eee{\end{eqnarray}}

\def \BC {\begin{cases}}
\def \EC {\end{cases}}

\begin{document}
\title{Aharonov-Bohm conductance of
a disordered single-channel quantum ring
  }
\author{ P.M. Shmakov, A.P. Dmitriev and V.Yu.Kachorovskii}
\affiliation{A.F.~Ioffe Physical-Technical Institute, 26 Polytechnicheskaya Street, St. Petersburg, 194021, Russia }

\begin{abstract}
We study the effect of {weak} disorder  on tunneling conductance of  a  single-channel quantum ring threaded by magnetic flux.
We assume that temperature is higher than the level spacing in the ring and  smaller than the Fermi energy.
In the absence of disorder, the conductance  shows sharp dips (antiresonances)
as a function of magnetic flux.
 We discuss different types of disorder and find that   the  short-range disorder    broadens antiresonances, while
  the long-range one leads to arising of additional resonant  dips.   We demonstrate that the    resonant dips have  essentially non-Lorentzian shape.
 The results are generalized to account for the spin-orbit interaction which leads to  splitting of the disorder-broadened  resonant dips,  and consequently  to coexisting of   two types of oscillations (both having  the form of sharp  dips): Aharonov-Bohm oscillations with magnetic flux and Aharonov-Casher oscillations with the strength of the spin-orbit coupling. We also discuss the effect of the  Zeeman coupling.
 \end{abstract}
\maketitle
\thispagestyle{plain}

\section*{Introduction}


The Aharonov-Bohm (AB) effect~\cite{bohmD}     is one of the   beautiful manifestations of the wave nature of  electrons.  The key physical issue --- the sensitivity of the  phase of an electronic wavefunction  to  a magnetic flux --- enables the  design
of quantum  AB  interferometers \cite{aronov87D,AB1,AB2,AB3,AB4,AB5,AB6,AB7,AB8,AB9,Preden1,Preden2,Bykov1,Bykov2,Zhang1,Ofek,Weisz,Yacoby5,Chang}   that can be tuned  by an external
magnetic field. Such interferometers occupy a worthy place in the quantum interferometry based on  low-dimensional electronic nanosystems.
A  {\it single-channel ballistic} ring tunnel-coupled to the leads and threaded  by the magnetic flux  is the
simplest realization of the AB interferometer (see Fig.~1).
The interference of  clockwise and counterclockwise  electron trajectories
manifests itself in the oscillations of the ring
conductance  $G(\phi)$
with the period $1$ (here $\phi$ is the magnetic flux measured in the units of the flux quantum $hc/e$). \cite{bohmD,aronov87D}

At low temperature $T$    and weak tunneling coupling, AB conductance exhibits
narrow
resonant peaks both in clean and disordered single-channel rings  \cite{buttD} (see also Refs.~\onlinecite{Moskalets1,Li, Mao, Feldman, Kokoreva} for discussion of  disordered  case). The peak arises
   each time when one of the field-dependent  energy levels in the ring crosses the Fermi energy $E_F.$
 Hence,  the positions of the AB resonances depend on $E_F$  \cite{buttD} (AB resonances are also affected by
the Coulomb blockade \cite{kinD,Grifoni}).
 Based on this physical picture one could expect the suppression of the  resonance structure at   $T\gg \Delta,$ where  $\Delta$ is the level spacing in the ring.
       Remarkably,  this naive expectation is incorrect and the interference effects are not entirely suppressed  by the thermal averaging.
Specifically,  for $T\gg \Delta$ the conductance of the  noninteracting  ring  with  weak tunnel coupling to the contacts exhibits sharp narrow dips (antiresonances) at $\phi = 1/2 +n,$  where  $n$ is an arbitrary integer number (see Fig.~2). \cite{jaglaD,dmitrievD}
It was also shown that
the electron-electron interaction  leads to
arising of a fine structure  of the antiresonances: each antiresonance splits
into a series of  narrow dips which correspond to blocking of the tunneling current by the  persistent one  \cite{dmitrievD}
(in contrast to the Coulomb blockade this effect is robust to increasing of temperature).

Additional physics comes into play in the presence of the  spin-orbit (SO) interaction.  In particular,
the rotation of the electron spin in the built-in SO  magnetic field  results in a spin phase shift  between clockwise and counterclockwise waves.
  This  phase
    is additional with respect to  AB phase and exists even at zero external magnetic field ($\phi=0$)
         so that  zero-field conductance exhibits   the  Aharonov-Casher (AC) effect: \cite{ACD,StoneD}
         periodic oscillations  with the strength of the SO coupling. The AC oscillations were the  focus of intensive  theoretical \cite{Stone1D,history1D,history6D,history2D,history3D,history4D,history5D,citro2D,pletD,citro1D,kovD,RomeoD,lobosD,plet1D,moldovD,AharonyD,MichettiD} research  and  their signatures  were  observed experimentally.\cite{exp1D,exp2D}
Recently, we demonstrated that these oscillations  are also not suppressed by thermal averaging, \cite{my} in a full analogy with the AB ones. Specifically, at $T\gg\Delta,$ SO interaction
splits   AB antiresonances into pairs of symmetrical (with respect to $\phi = 1/2 +n$)  antiresonances.  We also showed that the Zeeman interaction leads  to appearance of two
additional negative peaks on each period \cite{my}.

What, to the best of our knowledge,  has not  been discussed in the literature  is the  effect of {\it disorder} on the tunneling conductance through a {\it single-channel}  ring at relatively high temperatures, $T\gg\Delta$.\cite{Moskalets} The aim of the current study  is to fill this gap.

In this paper, we  study the  tunneling transport of non-interacting electrons through a disordered single-channel  quantum ring of length $L$ threaded by a
magnetic flux $\phi.$ We assume that $T$ is much smaller than the Fermi energy $E_F$ but large  compared to  $\Delta=2\pi\hbar v_F/L$
(throughout the paper  we linearize electron spectrum near $E_F,$ thus   neglecting small variation of the electron velocity   within the temperature band).
  The tunneling coupling  characterized by tunneling probability
 $\gamma$ is assumed  to be weak, $\gamma \ll 1,$  which implies that  the ring is almost closed.

   We discuss different types of disorder and find that the short-range  disorder    broadens antiresonances at $\phi= n+1/2$ while the long-range  one leads to arising of additional antiresonances  at $\phi=n.$
    We also find that the    resonant dips have  essentially non-Lorentzian shape.
 The results are generalized to account for the spin-orbit interaction which leads to  splitting of the disorder-broadened  resonant dips,  and consequently  to coexisting of   two types of oscillations (both having  the form of sharp periodic dips): Aharonov-Bohm oscillations with magnetic flux and Aharonov-Casher oscillations with the strength of the spin-orbit coupling. Additional disorder-broadened  resonant dips arise  in the presence of the  Zeeman coupling.

\section{ Clean ring} \label{clean}

    We start with discussion of  the  high-temperature conductance of the  clean ring following Refs.~\onlinecite{dmitrievD,my}.     This section aims  to introduce  basic notions and  clarify our approach to the problem. Later this approach  will be generalized to describe the effect of  disorder.

\begin{figure}[ht!]
 \vspace{3.5mm}
 \leavevmode \epsfxsize=5.0cm \centering{\epsfbox{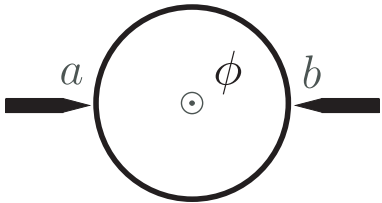}}
\caption{The ring threaded by magnetic flux $\phi.$
\vspace{3.5mm}
}
\label{fig1}
\end{figure}

The conductance is given by the Landauer formula:
\begin{equation}
G(\phi)=\frac{e^2}{\pi\hbar} {\cal T}(\phi),
\label{Landauer}
\end{equation}
where
\begin{equation}
{\cal T}(\phi)= \langle{\cal T}(\phi,E) \rangle_E =-\int {\cal T}(\phi,E) \frac{\partial f}{\partial E}dE,
\label{average}
\end{equation}
is the thermal average of the  transmission coefficient   ${\cal T}(\phi,E)$ and   $f(E)$ is the Fermi-Dirac function (here we take into account double spin degeneracy).

We  consider symmetrical setup  (see Fig.~\ref{fig1}) with identical  point contacts   described by the scattering matrix,
\begin{equation}
S = \begin{bmatrix} t_r & t_{out} & t_{out} \\ t & t_b & t_{in} \\ t & t_{in} & t_b \end{bmatrix},
\label{S}
\end{equation}
whose  elements \cite{buttD}
\begin{eqnarray}
t_{in} = \frac{1}{1+\gamma},\,\,\,t_b= -\frac{\gamma}{1+\gamma},\nonumber\\
t=t_{out} = \frac{\sqrt{2\gamma}}{1+\gamma},\,\,\, t_r = -\frac{1-\gamma}{1+\gamma}.
\label{t}
\end{eqnarray}
represent amplitudes  of scattering from  three incoming channels $(1,2,3)$  to three outgoing ones  $(1',2',3')$ (see Fig.~3).
Here $\gamma$ is  a real
parameter characterizing the strength of  the tunneling coupling to the  contact: weak  coupling corresponds  to $\gamma \ll 1, $ while  an open   contact is described by $\gamma \sim 1. $

\begin{figure}[ht!]
 \leavevmode \epsfxsize=5cm
 \centering{\epsfbox{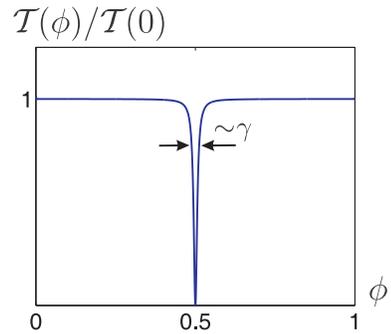}}
\caption{{ Antiresonance in high-temperature transmission coefficient in the absence of disorder.}
\vspace{-3.5mm}}
\label{fig2}
\end{figure}

 The transmission amplitude can be calculated by summation of the amplitudes of all the trajectories connecting contact $a$ and contact $b,$ including the trajectories with backscatterings by contacts (the processes $2\to 2'$ and $3 \to 3'$ on Fig. 3).
 Let us denote by $n$ the number of times the electron passes the contact $b,$ without exiting the ring ($n = 0, 1,\ldots$).
 The trajectories with a given $n$ consist of the odd number $2n+1$ of semicircles and thus have the same length $L_n = L(n+1/2)$. The sum of the amplitudes of
such trajectories  can be written as $\beta_n\exp(i k L_n),$ where $k = \sqrt{2mE}/\hbar$ is the electron wavenumber.  Hence, the transmission amplitude  is written as
\be
 t(\phi,E)=\sum_{n=0}^{\infty} \beta_n\exp(i k L_n),
 \label{amp}\ee
Next, we separate contributions of trajectories ending with lower and  upper semicircle thus writing  $\beta_n = \beta_n^+ + \beta_n^-.$
Introducing vector $\boldsymbol \beta_{n}$ with two components,    $  \beta_{n}^+$ and $ \beta_{n}^-,$   one may easily derive  the following recurrence relations:
\be
 \boldsymbol\beta_{n+1}= \hat A \boldsymbol \beta_{n},
\label{A0}
\ee
where
the  the  matrix $\hat A$ is given by
\bee
\label{A00}
 \hat A  &=& \begin{bmatrix} t_{in}^2 e^{-2\pi i\phi} +t_b^2 & t_b t_{in} (e^{-2\pi i\phi} +1)\\ t_b t_{in} (e^{2\pi i\phi} +1) & t_{in}^2 e^{2\pi i\phi} +t_b^2 \end{bmatrix} \\  \nonumber\\
   &=& \frac{1}{(1+\gamma)^2} \begin{bmatrix}  e^{-2\pi i\phi} +\gamma^2 &  -\gamma(e^{-2\pi i\phi} +1)\\ -\gamma (e^{2\pi i\phi} +1) &  e^{2\pi i\phi} +\gamma^2 \end{bmatrix} , \label{A000}
\eee
The element   $ A_{ij} $ [multiplied by $\exp(ikL)$] is the sum of the
amplitudes of  the trajectories     starting at the contact $b$
 and making a single return to the same contact (indices $i$ and $j$
 specify, respectively, the final and initial directions of motion: $i=\pm,j=\pm$).
The components of the vector $\boldsymbol \beta_0,$
\be \beta_{0}^+=t t_{out}e^{-i\pi\phi} , ~~~\beta_{0}^-=t t_{out}e^{i\pi\phi} \label{beta00}, \ee
 yield  contributions  of shortest counterclockwise and clockwise   trajectories, respectively.

\begin{figure}[ht!]
 \leavevmode \epsfxsize=2.5cm
 \centering{\epsfbox{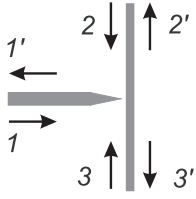}}
\caption{{ Scattering on contacts: the amplitude $t$ corresponds to processes $1\to 2'$ and $1 \to 3'$, $t_{out}$ - to $2\to 1'$ and $3\to 1'$, $t_r$ - to $1\to1'$,  $t_{in}$ - to $2\to 3'$ and $3 \to 2'$, $t_b$ - to $2\to 2'$ and $3 \to 3'.$}
\vspace{-3.5mm}}
\label{fig3}
\end{figure}

Using Eq.~\eqref{amp}, we express transmission coefficient  in terms of $\beta_n:$
\begin{equation}
{\cal T}(\phi,E) =  |t(\phi,E)|^2=\sum\limits_{n,m=0}^{\infty}\beta_n  \beta_m^* e^{ik(L_n-L_m)}.\label{double_sum}
\end{equation}
%
 The terms with $n\neq m$ in Eq.~\eqref{double_sum} vanish after thermal averaging in the discussed case  $T\gg \Delta,$ so that
the  expression for the averaged  transmission coefficient becomes
\be
{\cal T}(\phi) = \sum\limits_{n=0}^{\infty}|\beta_n|^2
= \sum\limits_{n=0}^{\infty}\,\left| ( \mathbf e, \hat A^n  \boldsymbol \beta_0 )\right|^2.\label{T0_sum2}
\ee
where vector $ \mathbf e$ has components $(1,1).$
%
%
%
%
%
%
The calculation of the sum entering  Eq.\eqref{T0_sum2} is quite cumbersome but straightforward (see  Appendix A). Using Eq.~\eqref{A**}
we obtain:\cite{dmitrievD}
\begin{equation}
{\cal T}(\phi) =\frac{2\gamma \cos^2\pi\phi}{\gamma^2+\cos^2\pi\phi}.\label{T0}
\end{equation}
The dependence ${\cal T}(\phi)$ is shown on Fig.~2.  The physical explanation of the  dip  (antiresonance) at $\phi = 1/2$
is quite simple (here and below we consider the interval $0<\phi<1$ ).  Let us demonstrate that at $\phi = 1/2$ the contribution of any  trajectory  is exactly canceled
by contribution
   of the trajectory mirrored  with respect to the  line connecting $a$ and $b$. Indeed, the sum of the amplitudes of these  two trajectories is proportional to $e^{ikL_n} (e^{i(2|m|+1)\pi\phi}+e^{-i(2|m|+1)\pi\phi})$, where $m$ is a difference between the number of  clockwise and counterclockwise revolutions, $|m|\leq n.$ At $\phi=1/2$ this sum turns to zero for any $k.$
Thus, the antiresonance    is due to the destructive interference of mirrored paths.

For weak tunneling coupling,  $\gamma\ll 1,$
the antiresonance is well approximated  by the  Lorentz-shape dip:
\begin{equation}
{\cal T}(\phi) \approx
2\gamma\frac{ \pi^2\delta\phi^2}{\gamma^2+\pi^2\delta\phi^2},\label{T00}
\end{equation}
where $\delta\phi  = \phi-1/2$.

 It is worth noting, that in the  vicinity of antiresonance  one can neglect  the backscattering on the contacts.
 Indeed, at   $|\phi -1/2| \sim \gamma $ the off-diagonal elements of $\hat A$
 which are proportional to $t_b \sim \gamma$  multiplied
 by  small factors $ 1+\exp(\pm i 2\pi\phi) \sim \gamma$ which implies that
 backscattering  is effectively suppressed by a factor $\gamma.$
 Physically, the effective suppression of backscattering   is explained  by destructive interference of two processes. In the first process an electron  is reflected  by a contact (say, contact $b$) and returns to this contact after one  revolution  around the ring.
  The amplitude of such a process is $t_b t_{in}\exp(\pm i2\pi\phi)$ (the sign  is prescribed by direction of the  propagation) where the amplitude $t_{in}$ appeared because the  contact $a$ was passed without reflection.  In the second process the electron passes the contact $b$ without reflection and then is reflected by contact $a$ and returns to $b.$ The corresponding amplitude is  given by $t_{in} t_b.$  Evidently, for $\phi=1/2$ the  amplitudes of these processes exactly cancel  each other.

  It is worth noting that backscattering is important in vicinity of integer values of flux.  \cite{dmitrievD} In particular,   putting $t_b=0$ in Eq.~\eqref{A00}  we come to incorrect conclusion that there are   resonant peaks at $\phi=n$ in the evident contradiction with Eq.~\eqref{T0}.


The approach discussed above allows one to find transmission coefficient for arbitrary $\gamma$ and $\phi.$  However,
it is technically cumbersome and lacks  physical transparency.
  Below we derive the main result of this section, Eq.~\eqref{T00}, by using an  alternative method.
  This method is valid only in the vicinity of $\phi=1/2, $ where backscattering by contacts  can be neglected. However, it  has a number of  advantages compared to the first one: it is more  illustrative physically, and much more easily generalized to account
for disorder.

 The key idea is that for $\gamma \ll 1$  and  $\delta \phi \ll 1$  the  tunneling amplitude through the ring may be presented as a sum of the transition amplitudes through  intermediate states corresponding to  quasistationary levels of almost closed ring.         The appropriate  analytical expression is derived in Appendix~\ref{CC} and reads
\be
{\cal T} \approx  \hbar^2 v_F^2 t^2t_{out}^2\left\langle\left|G_{E+i\Gamma/2}(0,L/2)\right|^2\right\rangle_E,\label{T0A*}
\ee
where
\be
G_{E+i\Gamma/2}(0,L/2) =\frac{1}{\hbar v_F } \sum_{l} C_{l} (E) =\sum_l\frac{\psi_l^*(0)\psi_l(L/2)}{E-\epsilon_l+i\Gamma/2},  \label{T0A2*}
\ee
 $G_{E}(0,L/2)$ is the Green function  of the closed ring,
  describing the transition from the contact $a$ to contact $b,$
$E$  is the energy of the tunneling electron, and $\epsilon_{l} $ are the electron energies in the closed ring, corresponding to 
 wave functions  $\psi _{l} (x).$  The quantities
\be C_{l}(E)=\hbar v_F\frac{\psi_{l}^{*}(0)\psi_{l}(L/2)}{E-\epsilon_{l}+i\Gamma/2} \label{C*}\ee
 are the amplitudes   of transition  through the corresponding quasistationary states (see Fig.~4).
  Here $\Gamma$ is the  tunneling rate
  given by
\be \Gamma = \frac{2\Delta \gamma}{\pi} .\label{Gamm*}\ee

In the closed clean ring there are two types of the electron states, corresponding to counterclockwise  and clockwise  propagation, labeled below by indices   $l=(n,+)$ and $l=(n,-)$ respectively.  Wave functions and energies of these states read
\be \psi_{n}^{\pm}(x) =  \frac{e^{\pm i  2\pi n x/L}}{\sqrt{L}},~ \epsilon_{n}^{\pm}(\phi) =\Delta(n \pm \phi) +c, \label{psi0*}\ee
where
constant $c=E_F-\Delta n_F$  arises in course of linearization of the spectrum near the Fermi
energy [here $n_F$ obeys the following equation: $E_F=2\hbar^2\pi^2 n_F^2/mL^2$].
As seen, for $\phi=1/2$ each level is double-degenerate, $ \epsilon_{n}^+(1/2)=\epsilon_{n+1}^-(1/2).$  Finite $\delta\phi$ lifts the degeneracy  of these levels:
\be
\epsilon_{n}^+=E_n +\Delta \delta\phi,~\epsilon_{n+1}^-=E_n -\Delta \delta\phi,
\ee
where $E_n=\Delta (n+1/2)+c.$

Let us now demonstrate that   Eqs.~\eqref{T0A*}, and \eqref{T0A2*}   yield ${\cal T}(1/2)=0.$
 Indeed,   from Eqs.~\eqref{C*}, and \eqref{psi0*}  we easily find:
 \be C_n^+ =-C_{n+1}^-, \hspace{3mm}\text{for} ~{\phi=1/2},\label{Ccancel}\ee
 so that $G_E(0,L/2)|_{\phi=1/2}  \equiv0$ for any $E$ and transmission coefficient turns to zero even before energy averaging.  The minus sign in the r.h.s. of Eq.~\eqref{Ccancel} appeared due to the property
 \be \psi_n^+ (L/2) =-\psi_{n+1}^-(L/2). \label{psipsi}\ee
This is the property that leads to  destructive interference
 and formation of the dip in the tunneling conductance.
Physically, this  is an alternative way to describe a compensation of  mirrored  paths   discussed above.

  Next we derive Eq.~\eqref{T00}.
   First, we rewrite \eqref{T0A*}  as follows
\begin{equation} \label{(5)}
{\cal T} =-t^{2} t_{out}^{2} \int \sum _{l,l'}C_{l}^{*}  (E)C_{l'} (E) \frac{\partial f}{\partial E} dE.
\end{equation}
The double sum in this equation contains  ``classical'' terms proportional to   $\left|C_{l} \right|^{2} $,  as well as the interference ones,   $C_{l}^{*} C_{l'} $ (with $l\ne l'$).   The main contribution to the integral in Eq.~\eqref{(5)} comes from  the vicinities of  poles (with the size on the order of  $\Gamma $) of the
amplitudes $C_{l}^{*} (E)$ and  $C_{l'} (E)$.  It is easy to see from Eqs.~\eqref{C*} and \eqref{(5)} that interference terms  are comparable with classical ones only if  $|\epsilon_l-\epsilon_{l^\prime}| \lesssim \Gamma.$   For $\delta \phi \ll 1$,
energies $\epsilon_{n}^{+} $ and $\epsilon_{n+1}^{-} $ are close to each other,
$\epsilon_{n}^{+} -\epsilon_{n+1}^{-} \approx 2\Delta \delta \phi $ (see Fig.~4),  and differ from the energies of other
 levels by  a much larger  distance  ($\Delta $ or larger).
The  interference contributions to Eq.~\eqref{(5)}   containing  products of  the amplitudes from different pairs, can be   neglected   compared to the ``classical'' terms  and to the interference  terms, containing products of the amplitudes from the same pair.  Therefore,  we can rewrite  Eq.~\eqref{(5)} as a sum over pairs of close levels
\begin{equation} \label{(6)}
{\cal T} \approx -t^{2} t_{out}^{2} \sum _{n}\int \left|C_{n}^{+} +C_{n+1}^{-} \right|^{2}   \frac{\partial f}{\partial E} dE.
\end{equation}
 Hence, the  paths over which electron passes through the ring can be split into pairs of interfering paths,  corresponding to quasi-degenerate intermediate states $\psi _{n}^{+} $ and $\psi _{n+1}^{-} $(see Fig.~4).

\begin{figure}[ht!]
  \leavevmode \epsfxsize=9.0cm
 \centering{\epsfbox{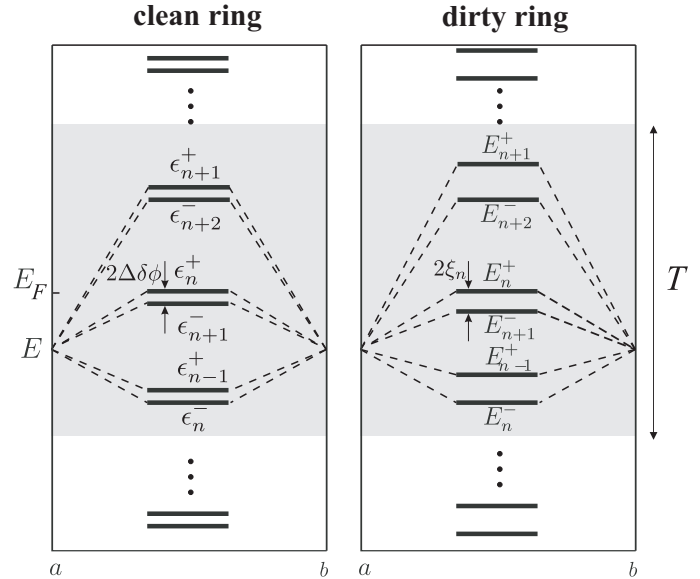}}
\caption{{ Tunneling of an electron through  pairs of close levels in the  ring. For  a clean ring the distance between levels in all pairs is the same and is given by $ 2\Delta \delta\phi.$ For a ring with disorder this distance  increases due to the repulsion of the levels in the disorder potential and becomes $n-$dependent: $E_n^+-E_{n+1}^-=2\xi_n=2\sqrt{\Delta^2 \delta\phi^2+ |V_n|^2}$.     }
\vspace{-3.5mm}}
\label{fig4}
\end{figure}

Now we demonstrate that  contributions of different pairs in Eq.~\eqref{(6)}  in fact differ by the thermal factor only, so that
 the problem can be reduced to  analysis of  the transition through a single pair. First we notice, that the energy dependence of the thermal factor ${\partial f}/{\partial E}$ is smooth
 and
 in the $n-$th term of the sum one may replace ${\partial f}/{\partial E}$ with $({\partial f}/{\partial E})|_{E=E_n }.$   
 Next we change the integration variable in this term: $E \to \epsilon=E-E_n. $ Then, dependence on $n$ remains only in the factors $\psi_n^{\pm*}(0)\psi_n^{\pm}(L/2)=(-1)^n/L.$  This dependence disappears after calculation of modulus squared in   Eq.~\eqref{(6)}.
 Finally,
 we calculate sum over $n,$
 $\sum_n\left(- {\partial f}/{\partial E}\right)|_{E=E_n}\approx 1/\Delta$
 (here we use inequality $T\gg\Delta$), and arrive to the following equation
\be
{\cal T} \hspace{-1mm} = \hspace{-1mm}\frac{ t^2t_{out}^2\Delta}{4\pi^2}\int_{-\infty}^{\infty} \hspace{-2mm}d \epsilon \left| \frac{1}{\epsilon-\epsilon^+ +i\Gamma/2}-\frac{1}{\epsilon-\epsilon^-+ i\Gamma/2}\right|^2
\label{T0A5**}
\ee
which expresses  ${\cal T}$ in terms of transition through a single
pair of close levels with energies $\epsilon^{\pm}=\pm \Delta\delta\phi .$ The  minus sign in front of the second fraction in Eq.~\eqref{T0A5**}
appeared due to the property \eqref{psipsi}.
One may separate in Eq.~\eqref{T0A5**}
"classical"  contribution, ${\cal T}_{{\rm cl}}(\phi)$ (integral form the sum of the squared amplitudes),
from the interference one,  ${\cal T}_{{\rm int}}(\phi).$   
Performing integration, we find ${\cal T}_{{\rm cl}}(\phi)=2\gamma$ and ${\cal T}_{{\rm int}}(\phi)=-2\gamma^3/(\gamma^2+\pi^2\delta \phi^2).$  Summing these terms we
 restore  Eq.~\eqref{T00}. We notice that, as expected,  interference  term  gives significant contribution only in the region of
small $\delta \phi:$ $\left|\delta \phi \right|\lesssim \gamma .$

\section {The ring with impurities}

In the above calculations we considered  the  case of the clean ring. Now we discuss the effect of disorder on the high-temperature conductance of the ring.

\subsection{Long-range disorder} One of the  realizations of the disorder  is a weak smooth random potential with the correlation length much exceeding the electron Fermi wavelength. In this case, backscattering by disorder
  is exponentially suppressed, so that  the potential only leads to the additional  phase shift
 between the right  and left-moving electron waves propagating from  contact $a$  to contact $b$ along upper and lower shoulder of interferometer, respectively (with zero winding number). We denote the disorder-induced phase difference between these two  waves as $\Psi(E)$.  Such an interferometer is evidently equivalent to the clean one   having two arms with the lengths $(L - a)/2$
and $(L + a)/2,$ where  $ a\approx \Psi(E_F)/k_F.$ The conductance of the latter  interferometer  was calculated in  Ref.~\onlinecite{dmitrievD}. From Eq.~(A3) of Ref.~\onlinecite{dmitrievD} we find
 \be\mathcal{T} (\phi)= { F[\sin (\pi \phi),\sin ( {\Psi}/{2})]+F[\cos (\pi \phi),\cos ( {\Psi}/{2})]
 }, \label{Tsmooth}
\ee
where \begin{equation}F(x,y)=2\gamma\frac{x^2 y^2}{x^2+\gamma^2 y^2}.\label{F}\end{equation}
This equation is valid provided that $T \left(d\Psi/dE\right)_{E=E_F} \ll 1.$
 As seen, for $\Psi \neq 0$  there are two dips in the conductance (at $\phi=1/2 $ and at $\phi=0$), the  widths and the depths  of the dips being oscillating functions of   $\Psi=\Psi(E_F)$ [in particular, ${\cal T}(0) = 2\gamma \cos^2(\Psi/2),~~ {\cal T}(1/2) = 2\gamma \sin^2(\Psi/2)$]. Hence, long-range disorder leads to appearance of the additional antiresonance in the conductance at $\phi=0$ and modifies the antiresonance near $\phi=1/2$.


\subsection{Short-range disorder}\label{short}
\subsubsection{Calculation of the transmission coefficient }
Another realization of disorder is the potential created by weak short-range impurities, randomly
distributed along the ring   with the concentration $n_i.$
Let us
characterize
the strength of disorder by  the  scattering
rate in the infinite wire calculated by the golden rule. For short-range potential, transport
 and quantum  scattering rates coincide and are given by ${1}/{\tau}=  {2|r|^2 v_F n_i},$ where  $r$ is the
reflection amplitude for a single impurity  ($|r| \ll 1$). Substituting
 in this equation  $n_i=N/L$ (here $N$ is the number of impurities in the  ring) we get
\be \frac{1}{\tau}=\frac{N|r|^2\Delta}{\pi \hbar }. \label{tau} \ee
We restrict ourselves to discussion of the ballistic case,\cite{local}
 $v_F\tau\gg L$, or, equivalently, $N|r|^2\ll1.$

\begin{figure}[ht!]
 \leavevmode \epsfxsize=5.0cm
 \centering{\epsfbox{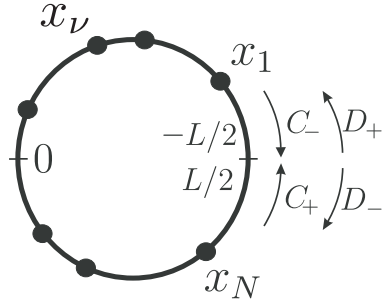}}
\caption{{A ring with impurities.}
\vspace{-3.5mm}}
\label{fig5}
\end{figure}

 We will see that the main effect of the short-range potential is the broadening of the antiresonances.  One could expect
 that scattering by disorder leads
 to essential increase of the  resonance width,  when $\tau$ becomes shorter than lifetime of the
 electron in the ring, $\hbar/\Gamma,$ which implies
  $N|r|^2\gg\gamma.$ Another expectation is that in the regime $N|r|^2\gg \gamma,$
 when electron experiences many scatterings during the lifetime and therefore acquires random phase, the interference is suppressed,
and, consequently, the depth of the dip essentially decreases.  However, we will show that
  the scattering on the impurities  comes into play at much smaller disorder strength when  $N|r|^2\sim \gamma^2,$  so that for $\gamma^2\ll N|r|^2\ll 1$ the dip is essentially broadened. Also,  in contrast to the naive expectation, its depth remains on the order of
   $\gamma.$


Now we generalize the method, introduced in the first part of the previous section. To do this we should
 modify the matrix $\hat A$, taking into account the scattering on the impurities.   This matrix
 becomes complicated, since it includes the amplitudes of all the trajectories with scatterings
 on both contacts and impurities, after which an electron returns to contact $b.$
 However, in the case $\delta\phi\ll1,$ $\gamma \ll1$ and $N|r|^2\ll1$ the matrix $\hat A$ can be  simplified.

 As a first step we expand matrix $\hat A$ in a Taylor series up to the first order   with respect to $\gamma ,$   denoting $\hat A_0= \hat A|_{\gamma=0}$.

 For the clean ring ($r=0$)  we have
\be
 \hat A
    \approx (1-2\gamma)\hat A_0\approx (1-2\gamma)  \begin{bmatrix}  e^{-2\pi i\phi}  &  0\\ 0 &  e^{2\pi i\phi}  \end{bmatrix} . \label{A0000}
    \ee
 We  neglected  off-diagonal elements of $\hat A,$ since, as we explained above, backscattering on the contacts is effectively suppressed at $\delta \phi \ll 1.$  Now we write the expansion for dirty ring in  a way  that reproduces Eq.~\eqref{A0000} for $r=0$
\be
\hat A= \hat A_0 -2\gamma (\hat A_0+\delta \hat A)+\ldots
\label{expansion}
\ee
Here both $\hat A_0$ and $\delta \hat A$  depend on $r.$ The matrix  $\delta \hat A$ should vanish at $r=0,$ so that $\delta \hat A \propto r$ at small $r.$ Calculations using Eq.~\eqref{A**} show
that  one can neglect the term $\gamma \delta \hat A$ in Eq.~\eqref{expansion} (as well as terms on the order of $\gamma^2$ and higher). Physically, this implies neglect of the processes involving  both scattering by impurities and forward scattering by contacts during one revolution around the ring.

 Let us now discuss the properties of the matrix $\hat A_0.$ According to the definition of the matrix $\hat A$ (see previous section) the matrix $e^{ikL}\hat A_0$ relates the amplitudes $C_\pm$ and $D_\pm$ of the incoming  and outcoming waves, respectively,  at the point $b$ (see Fig.~5)
\be
\begin{bmatrix} C_+ \\ C_- \end{bmatrix} = e^{ikL}\hat A_0 \begin{bmatrix} D_+ \\ D_- \end{bmatrix} \label{AS1}
\ee
Hence, the matrix $e^{ikL}\hat A_0$ is the $S-$matrix   describing   a complex scatterer consisting of   $N$ impurities, located at points $x_\nu$ between $x= -L/2$ and $x= L/2.$ Having in mind that
this matrix should be  unitary and taking into account
 the time reversal symmetry we  write this matrix in the most general form:
\be
\hat A_{0} = e^{i\alpha}\begin{bmatrix} \sqrt{1-|R|^2} e^{-2i\pi\phi } & R \\-R^*& \sqrt{1-|R|^2}e^{2i\pi\phi}\end{bmatrix}.\label{SS}
\ee
Here $\alpha$ is the small forward scattering phase for a complex scatterer consisting of     $N$ impurities.  This phase  is added to the geometrical phase $kL$ and, therefore, drops out after thermal averaging.
 The off-diagonal element,     $R,$  is,  up to a phase factor, the reflection amplitude
 from a complex of $N$ impurities.
 One can expand $R$ with respect to $r.$
In the lowest order in $r$  we obtain
\be
R\approx r \sum\limits_{\nu=1}^N e^{- 2 i k x_\nu}. \label{R}
\ee
This expression takes into account only one backscattering on impurities during a revolution around the ring, and is valid in the case $N|r|^2\ll1$.


As we see, the matrix $\hat A$ is now dependent on $k$, so that     Eq.\eqref{T0_sum2}  does not generally follow from Eq.\eqref{double_sum}. However, if we assume that
impurities are randomly distributed along the ring (some special non-random distributions will be discussed at the end of Sec.~\ref{short})
 the terms with $n\neq m$ in Eq.\eqref{double_sum} do not survive the averaging over $k$. Therefore we can use   Eq.\eqref{T0_sum2}  while performing  averaging over $k$:
\be
{\cal T} = \left\langle\sum\limits_{n=0}^{\infty}\,\left| ( \mathbf e, \hat A^n  \boldsymbol \beta_0)\right|^2\right\rangle_k. \label{TI_sum}
\ee
Since the probability to scatter on
an impurity  before the first visit of the point $b$  is small, we can neglect
the scattering terms in the vector $\boldsymbol\beta_0$  entering Eq.~\eqref{TI_sum} and use Eq.~\eqref{beta00}.
The sum in  Eq.~\eqref{TI_sum} can be calculated with the use of Eq.~\eqref{A**} which simplifies after expansion of the
numerator  and denominator with respect to $\gamma, r$ and $\delta \phi.$
 Calculations yield the following expression for the transmission coefficient:
\be
{\cal T}\approx 2\gamma \left\langle \frac{\pi^2\delta \phi^2 +|r|^2(\sum_\nu \sin 2k x_\nu)^2/4}{\pi^2\delta \phi^2+\gamma^2 +|r|^2|\sum_\nu e^{2ik x_\nu}|^2/4} \right\rangle_k.
\label{Txi}
\ee
This equation is valid provided that  $\gamma\ll1, N|r|^2\ll1$ and $\delta\phi\ll1.$
The relation between $\sqrt{N}|r|$ and $\gamma$ can be  arbitrary.

\begin{figure}[ht!]
 \leavevmode \epsfxsize=5.0cm
 \centering{\epsfbox{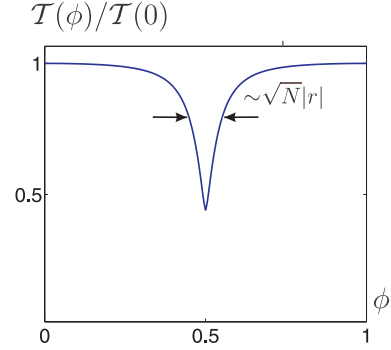}}
\caption{{Antiresonance in high-temperature transmission coefficient in the presence  of  $N$ randomly distributed impurities for $\sqrt{N}|r| \gg\gamma.$}
\vspace{-3.5mm}}
\label{fig6}
\end{figure}

The approach, discussed  above, can be also used to calculate the transmission coefficient for arbitrary $\phi.$ In particular,    one can show that at $\phi = 0$ there appears a small dip with the  amplitude on the order of $N|r|^2\gamma \ll \gamma$ (to obtain this result one should take into account backscattering by contacts).  In the following discussion we focus only on the antiresonance at $\phi = 1/2.$

Let us first consider a ring with a single impurity. As seen from Eq.~\eqref{Txi}, the transmission coefficient   is given by the Lorentz-shape antiresonance:
\begin{equation}
\mathcal{T} \approx 2\gamma\frac{\pi^2\delta\phi^2 + |r|^2/8}{\pi^2\delta\phi^2 +\gamma^2 + |r|^2/4}.
\label{single}\end{equation}
We see that the transmission coefficient at $\phi=1/2$ is no longer equal to zero and the antiresonance broadens so that
its width becomes $\sqrt{\gamma^2 + |r|^2/4}$. We also find that the  depth of the dip
changes from  $2\gamma$ to $\gamma$ with increasing $|r|.$  In other words, in contrast to the  antiresonance width, its depth  remains the same  order of magnitude.

In order to perform the averaging over $k$ in the case of many impurities, we notice, that
for random impurity distribution
the averaging over $k$ is equivalent to averaging over $x_\nu: \langle\cdots\rangle_k = \langle\cdots\rangle_{x_1\cdots x_N} $.
For two impurities the average  is easily calculated.   The result reads
\begin{equation}
\mathcal{T} \approx \gamma\left(\frac{\pi^2\delta\phi^2-\gamma^2}
{\sqrt{(\pi^2\delta\phi^2+\gamma^2)(\pi^2\delta\phi^2+\gamma^2+|r|^2)}}+1\right).
\label{T2}\end{equation}

 For the case $N>2$, we rewrite Eq.~\eqref{Txi} using the identities $x^{-1} \equiv \int_0^\infty \exp(-tx) dt$, $\exp(-x^2) \equiv \int_{-\infty}^{\infty} \exp(-y^2 +2ixy)dy/\sqrt{\pi},$ and get the following expression:
\bee
&&{\cal T}\approx 2\gamma \int\limits_0^{\infty}\frac{dt}{4\pi t} e^{-4(\pi^2\delta\phi^2+\gamma^2)t/|r|^2} \label{Tint}\\
&&\times \int\limits_{-\infty}^{\infty}d\xi d\eta e^{-\frac{\xi^2+\eta^2}{4t}}\left(\frac{4\pi^2\delta\phi^2}{|r|^2} -\frac{\partial^2}{\partial \xi^2}\right) J_0^N(\sqrt{\eta^2 +\xi^2}), \nonumber
\eee
or
\bee
&&{\cal T}\approx {2\gamma} \int\limits_{0}^{\infty} d\rho K_0\left(\frac{2\rho\sqrt{\pi^2\delta\phi^2+\gamma^2}}{|r|}\right)  \nonumber\\
&& \times \left(\frac{4\pi^2\delta\phi^2\rho}{|r|^2} -\frac{\partial}{2\partial \rho} \rho \frac{\partial}{\partial \rho}\right) J_0^N(\rho), \label{Tint2}
\eee
where $K_0$ is the modified Bessel function of the second kind.

Assuming now that the number of  impurities is large, $N\gg 1,$ we get  $J_0^N(x)\approx \exp(-Nx^2/4)$ and after simple calculation obtain
\begin{equation}
\mathcal{T} \approx \frac{2\gamma}{s^2}\int\limits_0^\infty \hspace{-1mm}dx \frac{\pi^2\delta\phi^2(1+x) + s^2/2}{(1+x)^2}\exp \hspace{-1mm}\left(\hspace{-1mm}-x\frac{\pi^2\delta\phi^2 +\gamma^2}{s^2}\right)\hspace{-1mm},
\label{largeN}
\end{equation}
where $s^2 = N|r|^2/4.$ This dependence  is plotted   in Fig.~6. We see that similar to the case of a single impurity, the transmission coefficient at $\phi=1/2$ is no longer equal
to zero and the antiresonance broadens. It is also notable that for $N\geqslant 2$ the dip has a non-Lorentzian shape.


Let us discuss  two limiting cases.  For $\sqrt N |r|\ll \gamma,$ the minimal value of conductance is given by $\mathcal{T}|_ {\delta\phi=0} \approx N |r|^2/4\gamma,$
and the width of  the antiresonances increases from $\gamma$ to $\gamma^\prime=\gamma+\delta$ where
\begin{equation} \delta \sim \frac{N |r|^2}{\gamma}
. \label{delta} \end{equation}
The relative  contribution  of the disorder to the resonance width, $\delta /\gamma \sim N|r|^2/\gamma^2= \hbar/\gamma\Gamma\tau,$   is enhanced  by a factor $\gamma^{-1} \gg 1$
in comparison with naive expectation $\hbar/ \Gamma \tau.$
In the opposite limiting case, $\sqrt N |r|\gg \gamma,$ we get $\mathcal{T}|_{\delta \phi=0}  \approx \gamma,$  so that the depth of the antiresonance is two times smaller compared with the case of clean ring, while the width is given by
\begin{equation} \gamma^\prime  \sim \sqrt{N} |r|.
 \label{gamma-prime}
 \end{equation}

 \subsubsection{Two-level approximation}\label{two}  Next we discuss the obtained results
  in terms of transition amplitudes through intermediate quasi-stationary states of almost closed ring.
  We recall that in the vicinity of the flux $\phi ={1/2} $  the energy levels of the clean ring can be split into pairs of close levels corresponding to clockwise and counterclockwise  propagation of the electron  inside the ring.  Energy distance between levels in the pairs equals to $2\Delta \delta \phi $ (see Fig.~4, left panel).  For the case when   impurity potential $V(x)=\sum _{\nu =1}^{N}U(x-x_{\nu } ) $ [here $U(x-x_{\nu } )$is the potential of the $\nu $-th impurity] is sufficiently weak (see corresponding criterion below) it can be simply accounted for by perturbation theory for two close levels.

   Calculation yields for  energies and wave functions of potential-disturbed states:
\bee
 E_n^+&=&E_n+ \xi_n ,~ E_{n+1}^-=E_n- \xi_n ,  \label{Enpm}\\
\xi_n&=&\sqrt{\Delta^2\delta\phi^2+|V_n|^2},\\
\label{xin}
%
 \Psi_n^+&=&\frac{\psi_n^+ ~+ \psi_{n+1}^-{V_n}/{W_n} }{\sqrt{1+|V_n|^2/W_n^2}},
\label{Psi1}
\\
 \Psi_{n+1}^-&=&\frac{\psi_{n+1}^- ~- \psi_n^+ V_n^*/W_n }{\sqrt{1+|V_n|^2/W_n^2}}
,\label{Psi2}
\eee
where
$W_n=\Delta\delta\phi + \xi_n,$
and
\be V_n =\int dx \psi_{n+1}^{-*}(x)V(x)\psi_n^+ (x)= \frac{i r\Delta }{2\pi } \sum_\nu e^{ 2\pi i(2n+1)x_\nu/L}.   \label{Vn}\ee
is the matrix element of the  impurity potential expressed in terms of reflection amplitude $r$.
For $\delta \phi \ll 1$, the inequality which ensures validity of the two-level perturbation theory   is given by $|V_{n} | \ll \Delta $.   For randomly distributed impurities, the amplitude of the potential is estimated as  $\left|V_{n} \right|\sim \left|r\right|\Delta \sqrt{N} $. Hence,  the perturbation theory applies  for $\left|r\right|\sqrt{N} \ll 1$.

 { Let us first discuss the dependence of the conductance on $\delta\phi$ qualitatively.
 First we note that due to the repulsion between levels in the impurity potential the minimal distance between levels in any pair is given by $|V_n|$
 (this distance corresponds to $\delta\phi=0$).
 %
    For
 $|V_n| \ll \Gamma $ similar to the case of a clean ring, there exist a dip in the transmission coefficient  which arises due to the
 interference  between  transition amplitudes  though pairs of close levels (see Fig.~4),
     while the "classical" term
     is featureless  at $\phi=1/2.$}

 {In contrast, in the opposite limiting case, $\left|V_{n} \right| \gg \Gamma, $
 the energy distance between levels in any pair becomes  much  larger than $ \Gamma $, and, consequently,  contribution of the interference terms
 to the transmission coefficient
 is small compared to ``classical'' ones.  This, however, does not lead to
 disappearance    of the dip in the transmission coefficient.  It turns out that in the dirty ring  the ``classical'' terms  acquire sharp dependence on
 $ \phi$ and decrease by a factor $2$ within a narrow region $\delta \phi \sim \sqrt{N}|r|.$}

 {Indeed, as seen from Eqs.~\eqref{xin}-\eqref{Psi2}, for $\left|\Delta \delta \phi \right|\gg \left|V_{n} \right|$  the wave functions in the $n-$th pair are simply given
 by clockwise- and counterclockwise-moving waves $\psi_n^+$ and $\psi_{n+1}^-.$ The ``classical'' contribution to the  transmission coefficient from each of these levels, say level $(n,+),$  is proportional to $|\psi_n^+(0)|^2|\psi_n^+(L/2)|^2=1/L^2.$ In the opposite case, $\left|\Delta \delta \phi \right|\ll \left|V_{n} \right|,$ disorder potential strongly mixes clockwise- and counterclockwise-propagating waves. Consider, for simplicity, the case   $\delta\phi=0.$  From Eqs.~\eqref{xin}-\eqref{Psi2} we see that $\Psi_n^+=(\psi_n^+ + e^{i\varphi_n} \psi_{n+1}^- )/\sqrt{2} $ and  $\Psi_{n+1}^-=(\psi_n^+ - e^{-i\varphi_n} \psi_{n+1}^- )/\sqrt{2}, $ where $\exp(i\varphi_n)=V_n/|V_n|.$  Averaging    $|\Psi_n^+(0)|^2|\Psi_n^+(L/2)|^2$  over random phase $\varphi_n$ and taking into account Eq.~\eqref{psipsi} we obtain twice smaller value  $1/2L^2.$ This implies existence of a dip
  of a width $\delta \phi  \sim |V_n|/\Delta \sim \sqrt{N}|r|$ in the transmission coefficient:  $\cal T$ decreases 
  by a factor $2$  within this width.} 

{The transition from the interference to ``classical'' picture of formation of the dip  can be illustrated  by an example of the ring with a single impurity. The conductance of such a ring is given by Eq.~\eqref{single}. The ``classical'' and  interference contributions to this expression read
 \bee
 \hspace{-3mm}{\cal T}_{cl}(\phi)\hspace{-2mm}&=&\hspace{-2mm} \frac{2\gamma(\pi^2\delta\phi^2+|r|^2/8)}{\pi^2\delta\phi^2+|r|^2/4},
 \\
 \hspace{-3mm}{\cal T}_{int}(\phi)&
 \hspace{-3mm}=&\hspace{-3mm}-\frac{2\gamma^3(\pi^2\delta\phi^2+|r|^2/8)}{(\pi^2\delta\phi^2+|r|^2/4)(\pi^2\delta\phi^2+\gamma^2
 +|r|^2/4)}.
 \eee
As seen,   the interference contribution  leads to formation of the dip at $\gamma \gg |r|$ and can be neglected for   $\gamma \ll |r|.$ }

{The rigorous calculations of the conductance can be performed in a way analogous to the case of the clean ring.} The transmission coefficient is expressed  via a Green function of the  closed ring and  can be  approximately presented  as
a sum over pairs of  interfering paths through  intermediate states $ \Psi _{n}^{+} $ and $ \Psi _{n+1}^{-} .$
Equations~\eqref{C*} and \eqref{(6)} still hold  with the following replacement:
 $\psi_n^\pm \to  \Psi_n^\pm $ and $\epsilon_n^\pm \to  E_n^\pm.$
%
In contrast to the case of the clean ring, the summands in Eq~\eqref{(6)} are now different  
not only due to the thermal factor $(\partial f/\partial E)|_{E=E_n}$  but also because of the strong dependence of $V_n$  on $n.$ However,  for random distribution of impurities
the summation over $n$ within the temperature window   is equivalent to averaging over the impurity positions. After such  averaging the dependence on $n$ remains only in the thermal factor and the problem   is again reduced to the case of   transition through  a single pair of close levels. The transmission coefficient may be written as a sum of classical and interference terms
\be
{\cal T}(\phi) = \frac{ t^2t_{out}^2}{\Delta}\int_{-\infty}^{\infty} d \epsilon \left(\rho_{{\rm cl}} + \rho_{{\rm int}}\right).
\label{T0A5***}
\ee
Here  $\rho_{{\rm cl}}=\langle | C_n^{+}|^2+ | C_{n+1}^{-}|^2 \rangle_{x_\nu}
,
~\rho_{{\rm int}}= 2{\rm Re} \langle  C_n^{+*} C_{n+1}^{-}\rangle_{x_\nu},$ where  $ C_n^{+},  C_{n+1}^{-}$ are now expressed via energy $\epsilon$ in the following way:
$ C_n^+=\hbar v_F\Psi_n^{+*}(0) \Psi_n^+(L/2)/(\epsilon-\xi_n +i\Gamma/2),   C_{n+1}^-=\hbar v_F\Psi_{n+1}^{*-}(0) \Psi_{n+1}^-(L/2)/(\epsilon+\xi_n+i\Gamma/2).$
Using Eqs.~\eqref{Enpm}-\eqref{T0A5***} after some algebra we arrive to Eq.~\eqref{Txi}.

\subsubsection{Special impurity distributions}
 Above we assumed that impurities are randomly distributed along the ring. If this is not the case,  the results might be quite different.
 The reason is that  there exist some impurity distributions  for which the summation over $n$ within the temperature band is not equivalent to the averaging over the impurity positions, so that calculation should be performed more carefully.  One may check that  the method discussed in Sec.~\ref{two} reproduces Eq.~\eqref{Txi} with the replacement
 $k \to k_n=2\pi(n+1/2)/L $  and  $\langle \cdots \rangle_k \to -\Delta \sum_n(\partial f/\partial E)_{k=k_n}\cdots.$
Let us now give some examples of special impurity distributions for which Eqs.~\eqref{T2}-\eqref{largeN} are invalid.  If  the impurities are distributed symmetrically with respect to the line $(a,b)$ connecting the contacts, then  $ \sum_\nu \sin 2k_n x_\nu=0$  and, consequently,    ${\cal T}(1/2) =0.$ On the language of  trajectories this can be explained by cancelation  of the contributions of the  mirrored paths just as in the case of the clean ring.  The width of the dip is on the order of $|r|\sqrt{N} $.  If impurities are distributed symmetrically with respect to the line perpendicular to $(a,b)$ and crossing the center of the ring, then $ \sum_\nu \cos 2k_n x_\nu=0$  and the  amplitude of the dip in ${\cal T} (\phi )$ goes to zero  with increasing $r$ for $\sqrt{N}|r| \gg \gamma.$
  Finally, if impurities are distributed symmetrically with respect to the ring center, one gets
  $\sum_{\nu}\exp(2ik_nx_\nu)=0$
 and dependence ${\cal T} (\phi )$ becomes the same as in the clean ring.

\section {The ring with impurities and spin-orbit interaction}
In this section we discuss the effect of impurities on the conductance of a ring with spin-orbit and Zeeman interactions. The case of a clean ring was studied in detail in Ref. \onlinecite{my}.

The Hamiltonian of a clean ring with SO-interaction induced by axially-symmetric built-in field is given by
\be
\hat H = \hat H_{kin} + \hat H_Z + \hat H_{SO},
\label{Ham}
\ee
where
\be
\hat H_{kin} = - \frac{\hbar^2}{2m}D_x^2,
\label{kin}
\ee
is the kinetic energy, $D_x = \partial/\partial x + 2\pi i\phi/L,$
\be
\hat H_Z = \frac12 \hbar \omega_Z\hat\sigma_z,
\label{ZM}
\ee
is the Zeeman term ($\hbar\omega_Z$ is the Zeeman splitting energy in the external magnetic field  parallel to the $z$ axis)   and  $\hat H_{SO}$ describes the SO coupling:
%
 \be
\hat H_{SO} = -i\xi\frac{\hbar^2}{2m}\left \{\begin{bmatrix} -\cos\theta && \sin\theta e^{-2\pi ix/L} \\ \sin\theta e^{2\pi ix/L} && \cos\theta \end{bmatrix}, D_x\right\}.
\label{SO}
 \ee
 Here
 $\theta$ is the angle between effective SO-induced magnetic field and the $z$ axis, $\xi$ is the dimensionless parameter  characterizing the strength of SO interaction, $\{ \ldots\}$ stands for the anticommutator.

 The problem is studied in the quasiclassical case ($k_F L \gg 1, \xi\ll k_F L$) in which the effect of the SO interaction is described by the rotation of the electron spin in effective magnetic field which varies along the electron trajectory. The stationary wavefunctions and energies in this case read:

 \bee
 &&\psi_{n,\pm}^{(1)}(x) = e^{\pm i 2 \pi n x/L}\begin{bmatrix} \cos\vartheta_\pm/2  \\ -\sin\vartheta_\pm/2\, e^{2\pi i x/L} \end{bmatrix},\label{psi_so}\\
&&\psi_{n,\pm}^{(2)}(x) = e^{\pm i 2 \pi n x/L}\begin{bmatrix} \sin\vartheta_\pm/2 \, e^{-2\pi i x/L}  \\ \cos\vartheta_\pm/2 \end{bmatrix},\nonumber\\
&&\epsilon_{n,+}^{(1)}=\Delta(n+\phi-\delta_+), ~~~ \epsilon_{n,-}^{(1)}=\Delta(n-\phi+\delta_-),\nonumber\\
&&\epsilon_{n,+}^{(2)}=\Delta(n+\phi+\delta_+), ~~~ \epsilon_{n,-}^{(2)}=\Delta(n-\phi-\delta_-).\nonumber
 \eee

Here we introduced the notations
\bee
&&\delta_\pm = |\varkappa_\pm| -\frac12, ~~~ e^{i\vartheta_\pm} = \frac{\varkappa_\pm}{|\varkappa_\pm|}, \label{kappa} \\
&&\varkappa_\pm =\frac12 + \xi e^{i\theta}  \mp\Omega_Z  , \nonumber\\
&&\Omega_Z = {\omega_Z L }/{4\pi v_F}. \nonumber
\eee

As seen from Eq.\eqref{psi_so} the degeneracy of the levels occurs for the following eight values of magnetic flux: $\phi = \pm\delta, \phi = \pm\delta^\prime,\phi = 1/2\pm\delta$ and $\phi = 1/2\pm\delta^\prime,$ where $\delta = (\delta_+ + \delta_-)/2, \delta^\prime = (\delta_+ - \delta_-)/2 +1/2.$ However, at four of these values, $\phi = \pm\delta$ and $\phi = \pm\delta^\prime$ the resonances are absent because  of  the backscattering on the contacts (see appendix B in Ref. \onlinecite{my}). At four other points, the backscattering is negligible, and there appear the antiresonances with the width $\gamma.$ The amplitudes of the antiresonances at $\phi = 1/2\pm\delta$ and
$\phi = 1/2\pm\delta^\prime$ are $\gamma c^2$ and $\gamma s^2,$ respectively, where $c = \cos(\vartheta_+-\vartheta_-)/2, s = \sin(\vartheta_+-\vartheta_-)/2.$

At $\phi = 1/2+\delta,$ the degeneracy occurs between the states $\psi_{n,+}^{(1)}$ and $\psi_{n+1,-}^{(1)};$ at $\phi = 1/2+\delta,$
between  $\psi_{n,+}^{(2)}$ and $\psi_{n+1,-}^{(2)};$ at $\phi = \delta^\prime-1/2,$  between  $\psi_{n,+}^{(1)}$ and $\psi_{n,-}^{(2)};$
and, finally, at $\phi = -\delta^\prime+1/2,$  between  $\psi_{n,+}^{(2)}$ and $\psi_{n,-}^{(1)}.$ We note that the amplitudes of the antiresonances ($\gamma c^2$ and $\gamma s^2$) are determined by the scalar products of the corresponding spinors: $c^2 = |\langle\psi_{n,+}^{(1)}|\psi_{n+1,-}^{(1)}\rangle|^2 = |\langle\psi_{n,+}^{(2)}|\psi_{n+1,-}^{(2)}\rangle|^2$ and $s^2 = |\langle\psi_{n,+}^{(2)}|\psi_{n,-}^{(1)}\rangle|^2 = |\langle\psi_{n,+}^{(1)}|\psi_{n,-}^{(2)}\rangle|^2.$  

The relation between the transmission coefficient and the stationary states of the closed ring, derived in Appendix B, allows us to easily find out the influence of the impurities on the four resonances, described above. As in the previous section, we use the two-level approximation (we assume that the distance between the antiresonances is much larger than $\sqrt{N}|r|).$ The matrix elements of impurity potential read:
 \bee
 &&\langle\psi_{n+}^{(1)}|\hat V|\psi_{m-}^{(1)}\rangle = \langle\psi_{n+}^{(2)}|\hat V|\psi_{m-}^{(2)}\rangle = c V_{nm},\label{v_so}\\
&&\langle\psi_{n+}^{(2)}|\hat V|\psi_{m-}^{(1)}\rangle = -\langle\psi_{n+}^{(1)}|\hat V|\psi_{m-}^{(2)}\rangle = s V_{nm}, \nonumber
 \eee
 where $V_{nm} = i\Delta r \sum_\nu \exp(i (n+m)2\pi x_\nu/L)/2\pi$ is the matrix element, appearing in the spinless problem. 

 Using Eqs.~\eqref{psi_so}, \eqref{v_so} and \eqref{T0A*} (the latter equation was modified for the spinful case) we obtain the following result for the transmission coefficient:

\bee
&&{\cal T}_{SO}(\phi) = \label{T_SOZ}  \frac{ c^2}{2}\left[{\cal T}(\phi +\delta; rc)+ {\cal T}(\phi -\delta;rc)\right] \\\nonumber &&+ \frac{s^2}{2}\left[{\cal T}(\phi +\delta^\prime;rs)+ {\cal T}(\phi -\delta^\prime;rs)\right],
\eee
where ${\cal T}(\phi; \rho)$ is the transmission coefficient in the spinless problem (given by Eq.\eqref{Txi}) with the reflection amplitude $r$ substituted by $\rho.$
We see that the effect of the impurity scattering in the spinless and in the spinful case is essentially the same: the antiresonances are broadened and their amplitudes become smaller (for strong enough impurities the amplitudes are two times smaller than in a clean ring). The only difference is the appearance of "effective" reflection amplitudes $rc$ and $rs$ for the antiresonances at $\phi = 1/2\pm \delta$ and $\phi = 1/2\pm \delta^\prime,$ respectively.

The expression for transmission coefficient is especially simple in the absence of Zeeman interaction (in this case $s = 0, c = 1$):
\be
{\cal T}_{SO}(\phi) = \frac12 \left[{\cal T}(\phi +\delta; r)+ {\cal T}(\phi -\delta;r)\right].
\ee
This equation shows two types of  periodic oscillations: AB oscillations with magnetic flux and AC oscillations with $\delta.$

\begin{figure}[ht!]
 \leavevmode \epsfxsize=5.0cm
 \centering{\epsfbox{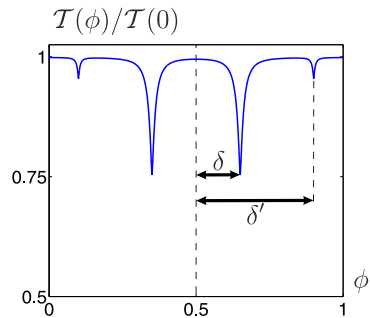}}
\caption{{ Transmission coefficient in the ring with impurities. Both spin-orbit and Zeeman interactions are present.}
\vspace{-3mm}}
\label{fig7}
\end{figure}

The  obtained results are illustrated in Fig.~\ref{fig7}.

\section{Conclusion}
In this work, we have studied  the effect of the disorder on the  Aharonov-Bohm interferometer made   of  a  single-channel non-interacting quantum ring  tunnel-coupled to the leads.  We focused on the case of large temperature (compared to the level spacing) and weak tunneling coupling. In this case tunneling conductance exhibits sharp dips at half-integer values of the magnetic flux.  We demonstrated that the short-range potential    broadens     these  dips, while
  the    long-range smooth disorder leads to appearing of   negative resonant peaks at integer values of the flux.
     We also found  analytical expression for the shape of the peaks which turned out to be  essentially non-Lorentzian.
 The results have been  generalized to account for the spin-orbit
 interaction which leads to splitting of the disorder-broadened  resonant peaks, and for the Zeeman coupling which results in arising of additional   peaks in the tunneling conductance.

\section{Acknowledgments}

We thank I.V. Gornyi and D.G. Polyakov for fruitful discussions and useful comments. The work was  supported by RFBR, by programmes of the RAS, by RF President Grant NSh-5442.2012.2, and by the Dynasty foundation.

\appendix
\section{}
In this appendix, we  calculate the sum
\be
\sum\limits_{n=0}^{\infty}\,\left|\left ( \boldsymbol \alpha,  \hat A^n \boldsymbol \beta \right) \right|^2,\label{A1}
\ee
where $\hat A$ is an arbitrary $2\times 2$ matrix, and
$\boldsymbol \alpha$ and $ \boldsymbol \beta$ are arbitrary two-component vectors.
First, we note that
\be
\hat A^n = \Sigma_n + \Delta_n (\hat A - \mathrm{Tr} \hat A/2), \label{A2}
\ee
where
$
\Sigma_n = {(\lambda_1^n+\lambda_2^n)}/2,\qquad \Delta_n = {(\lambda_1^n-\lambda_2^n)}/{(\lambda_1-\lambda_2)}.
$
 To prove Eq. \eqref{A2}, one can make a similarity transformation that reduces  $\hat A$  to a diagonal (or Jordan) form.
Using Eq. \eqref{A2} we can rewrite Eq.~\eqref{A1} as follows:
\bee
\nonumber
&&\sum\limits_{n=0}^{\infty}\,\left|\left ( \boldsymbol \alpha,  \hat A^n \boldsymbol \beta \right) \right|^2= \\ && \nonumber
|a|^2 \sum\limits_{n=0}^{\infty} |\Sigma_n|^2 +  |b - a \mathrm{Tr} \hat A/2|^2 \sum\limits_{n=0}^{\infty} |\Delta_n|^2 \\
&&+ 2\mathrm{Re}[a^*(b-a \mathrm{Tr} \hat A/2) \sum\limits_{n=0}^{\infty} \Sigma_n^*\Delta_n]\label{A4}
,
\eee
where
$ a = ( \boldsymbol \alpha ,  \boldsymbol \beta ), ~b=  ( \boldsymbol  \alpha,  \hat A  \boldsymbol \beta ).
$
The sums entering Eq.~\eqref{A4} are easily calculated:
\bee
&& \sum\limits_{n=0}^{\infty} |\Sigma_n|^2 =\nonumber \\&& \frac14\left(\frac1{1-|\lambda_1|^2}+\frac1{1-|\lambda_2|^2} +2 \rm{Re}\frac1{1-\lambda_1^*\lambda_2}\right),\nonumber\\
&& \sum\limits_{n=0}^{\infty} |\Delta_n|^2= \nonumber \\&&  \frac1{|\lambda_1-\lambda_2|^2}\left(\frac1{1-|\lambda_1|^2}+\frac1{1-|\lambda_2|^2} -2 \rm{Re}\frac1{1-\lambda_1^*\lambda_2}\right),\nonumber\\
&& \sum\limits_{n=0}^{\infty}  \Sigma_n^*\Delta_n = \nonumber \\&& \frac1{2(\lambda_1-\lambda_2)}\left(\frac1{1-|\lambda_1|^2}-\frac1{1-|\lambda_2|^2} +2i \rm{Im}\frac1{1-\lambda_2^*\lambda_1}\right)\nonumber
\eee
$$$$
These expressions can be rewritten in terms of $D = \mathrm{det} \hat A,~~  S = \mathrm{Tr}\, \hat A.$ After some algebra we obtain
\bee
&& \sum\limits_{n=0}^{\infty} |\Sigma_n|^2 =\nonumber \\&&  \frac14 Z^{-1}[4 (1-|D|^2)-|S|^2(|D|^2+3)+ 2 \mathrm{Re}\, D^*S^2]\nonumber\\
&& \sum\limits_{n=0}^{\infty} |\Delta_n|^2 = Z^{-1}(1-|D|^2),\nonumber\\
&& \sum\limits_{n=0}^{\infty}  \Sigma_n^*\Delta_n= \frac12 Z^{-1}[S^*(1+|D|^2)-2D^* S],
\eee
where $ Z = (1-|D|^2)^2-(1+|D|^2)|S|^2 + 2 \mathrm{Re} D^* S^2. $
Finally, we get
\widetext
\be
\sum\limits_{n=0}^{\infty}\,\left|\left ( \boldsymbol \alpha,  \hat A^n \boldsymbol \beta \right) \right|^2=
\frac{|a|^2(1-|D|^2 - |S|^2 - |DS|^2+ 2 \mathrm{Re} D^* S^2) + |b|^2(1-|D|^2) + 2\mathrm{Re}[a^*b(S^*|D|^2-D^*S)]}
{(1-|D|^2)^2-(1+|D|^2)|S|^2 + 2 \mathrm{Re} D^* S^2}.\label{A**}
\ee
In the case of real $D$ and $S$ this expression is simplified:
\be
\sum\limits_{n=0}^{\infty}\,\left|\left ( \boldsymbol \alpha,  \hat A^n \boldsymbol \beta \right) \right|^2 =
\frac{|a|^2(1+D - S^2 +DS^2) + |b|^2(1+D) - 2DS\mathrm{Re}(a^*b)}
{(1-D)[(1+D)^2-S^2]}.\label{A***}
\ee
\endwidetext

\section{}\label{CC}

In this Appendix we show that the transmission coefficient may be easily expressed  in terms of stationary levels of the closed ring. We start from discussion of the clean ring and then generalize obtained results for the ring with short-range disorder.
\subsection{Clean ring}\label{appclean}
Using Eqs.~\eqref{average}, \eqref{A0} and \eqref{double_sum}, one can  write
the transmission coefficient  in the following way
 \be
{\cal T}(\phi) =
\left\langle \left|\left ( \mathbf e,  ~ \frac{1}{1-e^{ikL}\hat A}~ \boldsymbol \beta_0  \right ) \right|^2 \right \rangle_{E}  , \label{T0_sum3}
\ee
   Neglecting the backscattering by the contacts [see equation  \eqref{A0000}]   we get
 \bee
 \label{T0_sum33}
{\cal T}(\phi) =
  t^2 t_{out}^2 \left \langle
 \left| \frac{e^{-i\pi\phi}}{1-(1-2\gamma)e^{i(kL-2\pi\phi)}} \right. \right.
\\\left.\left. +\frac{e^{i\pi\phi}}{1-(1-2\gamma)e^{i(kL+2\pi\phi)}}~  \right | ^2  \right \rangle_E ,
\nonumber
\eee
 For $\gamma =0,$   the integrand of Eq.~\eqref{T0_sum3}  has poles on the real axis,
    $k_{n}^{\pm}={2\pi}(n \pm\phi)/L.
    $
   These poles   are related to the energy levels  of the closed ring in following way [see Eq.~\eqref{psi0*}]: $ \epsilon_n^\pm=\hbar v_F k_n^\pm+c$
   (it is notable that $k_{n}^{\pm}$ do not coincide with the eigenvalues of momentum operator).
   Expanding denominators of these fractions near $k_{n}^{\pm}$, after simple algebra we
   find that Eq.~\eqref{T0_sum33} approximately coincides  with Eq.~\eqref{T0A*} of the
   main text where $G_E$ is found from Eq.~\eqref{T0A2*}.  In this case, the wave functions
   entering  Eq.~\eqref{T0A2*}  are simply given by  $\psi_n^{\pm}(0)=1/\sqrt{L}$ and
   $ \psi_n^{\pm}(L/2)=(-1)^n/\sqrt{L}.$ Below we demonstrate that  Eq.~\eqref{T0A*}
   also holds for dirty ring where wave functions strongly depend on realization of disorder.
\subsection{Ring with short-range disorder}\label{appdisorder}
 The transmission coefficient of disordered ring is also   given by   Eq.~\eqref{T0_sum3} where both $\hat A$
 and $ \boldsymbol \beta_0$ depend on disorder.  As we discussed in Sec.~\ref{short} the
  matrix $\hat A$ can be approximately written as $\hat A= (1-2\gamma) \hat A_0,$ where the unitary matrix $\hat A_0$ is given by Eq.~\eqref{SS}.
  Let us also introduce matrix $\hat U,$  such that $e^{ikL/2}\hat U$ is a transfer matrix from contact $a$ to contact $b.$ Then, one can express $ \boldsymbol \beta_0$ in terms of this matrix    $
  \boldsymbol \beta_0 = e^{ikL/2} t t_{out} \hat U  \boldsymbol e.
$ 
 Denoting eigenvalues  of $e^{ikL} \hat A_0$ as  $e^{iQ_1(k)L}$ and $ e^{iQ_2(k)L}$  and corresponding eigenvectors as  $ \boldsymbol \chi_1(k)$ and $ \boldsymbol \chi_2(k),$ we rewrite Eq.~\eqref{T0_sum3} as follows
  \be
%
{\cal T}(\phi) =t^2t_{out}^2\left\langle \left|\sum_{\alpha=1,2}
\frac{ ( \mathbf e,\boldsymbol \chi_\alpha) ( \boldsymbol \chi_\alpha,\hat U  \boldsymbol e ) }
{1-(1-2\gamma)e^{iQ_\alpha(k)L}}
\right|^2 \right \rangle_{E} .
\label{T0_sum3dirty}
\ee
%
%
%
Equation \eqref{T0_sum3dirty} is a generalization of Eq.~\eqref{T0_sum33} for a disordered ring. [
 In the  clean ring,   $Q_1(k)=k-2\pi\phi/L, Q_2(k)=k+2\pi\phi/L,$
  $~\boldsymbol  \chi_1 =(1,0),$
   $\boldsymbol \chi_2= (0,1),$ and $\hat U$ is diagonal matrix with the elements $\exp(-i\pi \phi)$ and  $\exp(i\pi \phi).$]

  As was  pointed out in the section \ref{short},  the matrix $\hat A$ relates the amplitudes $C_\pm, D_\pm$ of the waves in the vicinity of the contact $b$ to each other [see Fig.~5 and Eq. \eqref{AS1}].
 In the closed ring ($\gamma = 0$) the stationary states can be found from the conditions $C_+ = D_+$ and $C_- = D_-.$ This allows us to establish a relation between the matrix $\hat A_0$ and the stationary states of the closed ring. Specifically,
  the wavevectors $k = k_{1 n}$ and $k = k_{2 n}$, for which one of the eigenvalues of the matrix $e^{ikL} \hat A_0$  equals
   to unity, correspond to the energy levels.
  These wavevectors are found from the equation:
 $$
 Q_\alpha(k_{\alpha n}) =2\pi n/L
$$
   Each of the corresponding   eigenvectors  with components $(C_+, C_-)$  describes the stationary wavefunction in the vicinity of the point $b$ ($x$ close to  $L/2$):
 \bee &&\psi(x)=\nonumber \\&& \nonumber  \frac{C_+ e^{i(k_{\alpha n}-2\pi\phi/L)(x-L/2)} +C_- e^{-i (k_{\alpha n}+2\pi\phi/L) (x-L/2)}}{\sqrt{L}}.\eee
     As expected, for  $\gamma=0$ Eq.~\eqref{T0_sum3dirty} has poles as a function of $k$  for  $k= k_{1n}$ and $k = k_{2n}.$

  For almost closed ring
   the poles of the r.h.s. of Eq.~\eqref{T0_sum3dirty} slightly shift away from the  axis of real $k.$      
   Just as in the clean ring, the main contribution to the integral over $E$ comes from the poles of   the fractions $[ 1-(1-2\gamma)e^{iQ_\alpha (k)L}]^{-1},$ while the terms    $( \mathbf e, \boldsymbol \chi_{\alpha}) ( \boldsymbol \chi_{\alpha} , \hat U \boldsymbol e)$ as well as function $\partial f/\partial E$ can be taken at the poles for $\gamma=0.$
 Next, we  notice that the  vectors $\boldsymbol \chi_{\alpha}$  and  the matrix $\hat U$ are defined in such a way that
    \bee ( \boldsymbol e, \boldsymbol \chi_{\alpha}) |_{k=k_{\alpha n}} \nonumber  &=&\sqrt{L} ~\psi_{\alpha n} (L/2), \\ ( \boldsymbol  \chi_{\alpha} , \hat U \boldsymbol e)|_{k=k_{\alpha n}}   &=& \sqrt{L}~\psi_{\alpha n}^* (0), \nonumber\eee
 where $\psi_{\alpha n}(x)$ are the stationary wave functions of the ring with disorder.
 Using these equations, expanding denominators in Eq.~\eqref{T0_sum3dirty} near
 the poles and neglecting $dQ_{\alpha}/dk -1 \sim \sqrt{N} |r| \ll 1  $ with respect to unity,
 we arrive to Eqs.~\eqref{T0A*} and \eqref{T0A2*}.


\begin{thebibliography}{8}

\bibitem{bohmD} Y.~Aharonov,  D.~Bohm,  Phys. \ Rev. \ B {\bf 115}, 485 (1959).
\bibitem{aronov87D} A.G.~Aronov and Yu.V.~Sharvin, Rev.\ Mod.\ Phys.\ {\bf 59}, 755 (1987).



\bibitem{AB1} A. Yacoby,  M. Heiblum, D. Mahalu, and  H. Shtrikman,
 Phys. Rev. Lett. {\bf 74}, 4047 (1995).

\bibitem{Yacoby5}A. Yacoby, R. Schuster, and M. Heiblum
Phys. Rev. B {\bf 53}, 9583 (1996)


\bibitem{AB2} A. van Oudenaarden,  M. H. Devoret,  Yu.V. Nazarov, and
J. E. Mooij, Nature {\bf 391}, 768 (1998).

\bibitem{Bykov1} A. A. Bykov, A. K. Bakarov, L. V. Litvin, and A. I. Toropov, JETP Letters  {\bf 72}, 209 (2000).

\bibitem{Bykov2} A. A. Bykov, D. G. Baksheev, L. V. Litvin, V. P. Migal’, E. B. Ol’shanetskii, M. Casse', D. K. Maude, and J. C. Portal, JETP Letters, {\bf71}, 434 (2000).


\bibitem{AB3}  O. M. Auslaender,  A. Yacoby, R.  de Picciotto,  K. W. Baldwin,
L. N. Pfeiffer, and West K W,
Science {\bf 295}, 825 (2002).


    	
\bibitem{AB4}    Yang Ji,
    Yunchul Chung,
    D. Sprinzak,
    M. Heiblum,
    D. Mahalu, Hadas Shtrikman,
Nature {\bf 422}, 415 (2003)

\bibitem{AB5}P. Samuelsson,  E. V. Sukhorukov,  M. Buttiker,  Phys. Rev. Lett. {\bf 92}, 026805 (2004).
    	
 \bibitem{AB6}   M. Avinun-Kalish,
    M. Heiblum,
    O. Zarchin,
    D. Mahalu,
    V. Umansky,
Nature {\bf 436}, 529 (2005).

 \bibitem{AB7} I. Neder, M. Heiblum, Y. Levinson, D. Mahalu, and V. Umansky, Phys. Rev. Lett. {\bf 96}, 016804 (2006).   	
   \bibitem{AB8} I. Neder,
    N. Ofek,
    Y. Chung,
    M. Heiblum,
    D. Mahalu, V. Umansky, Nature {\bf 448}, 333 (2007).





\bibitem{AB9} I. Neder, M. Heiblum, D. Mahalu, and V. Umansky, Phys. Rev. Lett. {\bf 98}, 036803 (2007)

\bibitem{Preden1} Preden Roulleau, F. Portier, D. C. Glattli, and P. Roche, A. Cavanna, G. Faini, U. Gennser, and D. Mailly, Phys Rev B {\bf 76}, 161309 (2007).
\bibitem{Preden2} Preden Roulleau, F. Portier, and P. Roche Phys. Rev. Lett. {\bf 100}, 126802 (2008).




\bibitem{Chang} Dong-In Chang, Gyong Luck Khym, Kicheon Kang, Yunchul Chung, Hu-Jong Lee, Minky Seo, Moty Heiblum, Diana Mahalu, Vladimir Umansky
Nature Physics 4, 205 (2008).



\bibitem{Zhang1} Yiming Zhang, D. T. McClure, E. M. Levenson-Falk, and C. M. Marcus, L. N. Pfeiffer and K. W. West, Phys. Rev. B {\bf 79}, 241304 (2009).

\bibitem{Ofek}  N. Ofek, Aveek Bid, M. Heiblum, Ady Stern, V. Umansky, and D. Mahalu
PNAS 107, 5276 (2010).

\bibitem{Weisz} E. Weisz, H. K. Choi, M. Heiblum, Yuval Gefen, V. Umansky, and D. Mahalu
Phy. Rev. Lett. 109, 250401 (2012).




%
%
%




\bibitem{buttD} M.~B\"uttiker, Y.~Imry, and M.Ya.~Azbel, Phys.\ Rev.\ A {\bf 30}, 1982 (1984); Y.~Gefen, Y.~Imry, and M.Ya.~Azbel, Phys.\ Rev.\ Lett.\ {\bf 52}, 129 (1984); M.~B\"uttiker,  Y.~Imry, R.~Landauer,  S.~Pinhas, Phys. \ Rev.\ B {\bf 31}, 6207 (1985).



\bibitem{Moskalets1} M.V.~Moskalets, Low\ Temp.\ Phys. {\bf 23}, 824 (1997).

\bibitem{Li} Qiming Li and C. M. Soukoulis, Phys. Rev. B {\bf 33}, 7318 (1986).





\bibitem{Mao} J. M. Mao, Y. Huang, and J. M. Zhou, J. Appl. Phys. {\bf 73}, 1853 (1993).


\bibitem{Feldman} E.P. Nakhmedov, H.Feldmann, and R. Oppermann, Eur. Phys. J. B {\bf 16}, 515 (2000).




\bibitem{Kokoreva} M.A. Kokoreva, V.A. Margulis, M.A. Pyataev, Physica E,  {\bf 43},  1610 (2011).




\bibitem{kinD} J.M.~Kinaret, M.~Jonson, R.I.~Shekhter, S.~Eggert, Phys. \ Rev.\ B {\bf  57}, 3777 (1998).
\bibitem{Grifoni} M. Eroms, L. Mayrhofer, and M. Grifoni,
Phys. Rev. B {\bf 78}
075403 (2008).

\bibitem{jaglaD} E.A.~Jagla, C.A.~Balseiro, Phys.\ Rev.\ Lett. {\bf 70}, 639
(1993).

\bibitem{dmitrievD} A.P.~Dmitriev, I.V.~Gornyi, V.Yu.~Kachorovskii, D.G.~Polyakov
 Phys. \ Rev. \ Lett., {\bf 105}, 036402 (2010).



\bibitem{ACD} Y.~Aharonov, A.~Casher, Phys.\ Rev.\ Lett {\bf 53}, 319 (1984)

\bibitem{StoneD} H.~Mathur, A.D.~Stone, Phys.\ Rev.\ Lett {\bf 68}, 2964 (1992)

\bibitem{Stone1D} H.~Mathur, A.D.~Stone,  Phys.\ Rev.\ B {\bf 44}, 10957 (1991).

\bibitem{history1D} A.G.~Aronov, Y.B.~Lyanda-Geller, Phys.\ Rev.\ Lett {\bf 70}, 343 (1993).

\bibitem{history6D} T.Z.~Qian,  Z.B.~Su, Phys.\ Rev.\ Lett.\ {\bf 72}, 2311 (1994).

\bibitem{history2D}J.~Nitta, F.E.~Meijer,  H.~Takayanji, Appl.\ Phys.\ Lett {\bf 75}, 695 (1999).

\bibitem{history3D} D.~ Frustaglia, K.~ Richter, Phys.\ Rev.\ B {\bf 69}, 235310 (2004).

\bibitem{history4D} B.~ Molnar, F.M.~ Peeters, P.~ Vasilopoulos, Phys.\ Rev.\ B {\bf 69}, 155335 (2004).
\bibitem{history5D} U.~Aeberhard, K.~Wakabayashi, M.~Sigrist, Phys.\ Rev.\ B {\bf 72}, 075328 (2005).
\bibitem{citro2D} R.~Citro, F.~Romeo,  Phys.\ Rev.\ B {\bf 73}, 233304 (2006).

\bibitem{pletD} M.~Pletyukhov, V.~Gritsev,  N.~Pauget, Phys.\ Rev.\ B {\bf 74},
045301 (2006).

\bibitem{citro1D} R.~Citro, F.~Romeo, M.~Marinaro, Phys.\ Rev.\ B {\bf 74}, 115329 (2006).

\bibitem{kovD} A.A.~Kovalev, M.F.~Borunda, T.~Jungwirth, L.W.~Molenkamp, J.~Sinova, Phys. \ Rev. B {\bf 76}, 125307 (2007).

\bibitem{RomeoD} F.~Romeo, R.~Citro,   M.~Marinaro,  Phys.\ Rev.\ B {\bf 78}, 245309 (2008).

\bibitem{lobosD} A.M.~Lobos and A.A.~Aligia, Phys.\ Rev.\ Lett {\bf 100}, 016803 (2008).

\bibitem{plet1D} M.~Pletyukhov and U.~Z\"ulicke, Phys.\  Rev.\  B {\bf 77}, 193304 (2008).

\bibitem{moldovD} V.~Moldoveanu and B.~Tanatar, Phys. Rev. B {\bf  81}, 035326 (2010).

\bibitem{AharonyD} A.~Aharony, Y.~Tokura, G.Z.~Cohen, O.~Entin-Wohlman,  S.~Katsumoto, Phys.\ Rev.\ B {\bf 84}, 035323 (2011).

\bibitem{MichettiD} P.~Michetti and P.~Recher, Phys. Rev. B {\bf 83}, 125420 (2011).

\bibitem{exp1D} M.~Konig, A.~Tschetschetkin, E.M.~Hankiewicz, J.~Sinova, V.~Hock, V.~Daumer, M.~Schafer, C.R.~Becker, H.~Buhmann,
 L.W.~Molenkamp, Phys.\ Rev.\ Lett. {\bf 96}, 076804 (2006).

\bibitem{exp2D}T.~Bergsten, T.~Kobayashi, Y.~Sekine,  J.~Nitta, Phys.\ Rev.\ Lett. {\bf 97}, 196803 (2006).


\bibitem{my} P. M. Shmakov, A. P. Dmitriev, and V. Yu. Kachorovskii
Phys. Rev. B {\bf 85}, 075422 (2012)
 \bibitem{Moskalets} The AB conductance through the single-channel       ring    with a scattering potential barrier   in one of the arms   was discussed in Ref.~\onlinecite{Moskalets1} for the case of almost transparent contacts ($\gamma \simeq 1$). It was shown that interference part of the conductance is  not entirely suppressed for $T \gg \Delta .$

\bibitem{local} In this case the localization effects are absent.













%
%
%
%
%
%
%
%
%
%
%
%
%
%
%
%
%
%
%
%
%
%
%
%
%
%
%
%
%
%
%
%
%
%
%
%
%
%
%
%
%
%
%
%
%
%
%
%
%
%
%
%
%
%
%
%
%
%
%
%
%
%
%
%
%
%
%
%
%
%
%
%
%
%
%
%
%
%
%
%
%





\end{thebibliography}
\end{document}